\documentclass[oribibl]{llncs}

\usepackage{amsmath,amsfonts,amssymb,stmaryrd,cmll}
\usepackage{graphicx}
\usepackage[nohug,midshaft,tight]{diagrams}
\usepackage{jbw-bib-settings}

\usepackage{tikz,tikz-inet,net}

\DeclareMathOperator{\ob}{ob}

\newarrow{Eqto}=====

\newcommand{\bigcat}[1]{\mathsf{#1}}

\newcommand{\fonc}[1]{\mathsf{#1}}

\newcommand{\Formulae}{\mathcal{F}}
\newcommand{\commentthis}[1]{}
\renewcommand{\to}[3][.5cm]{%
  \begin{diagram}[inline,width=#1,tight] %
    #2 & \rTo & #3
  \end{diagram} %
}

\newcommand{\bsig}{\mathcal{K}} 
\newcommand{\bbig}{M} 
\newcommand{\T}{\fonc{T}} 
\newcommand{\theory}{\mathcal{T}}
\newcommand{\sig}{\Sigma}
\newcommand{\A}{\mathcal{A}}
\newcommand{\C}{\mathcal{C}}

\newcommand{\X}{\mathcal{X}}

\newcommand{\Sig}{\bigcat{SMCSig}} 

\newcommand{\SMCCat}{\bigcat{SMCCat}}

\newcommand{\piget}{\ensuremath{\mathit{g}}}
\newcommand{\pisend}{\ensuremath{\mathit{s}}}
\newcommand{\trou}{\boxempty}

\newcommand{\smc}{\textsc{smc}}
\newcommand{\dr}{\textsc{dr}}
\newcommand{\hoas}{\textsc{hoas}}
\newcommand{\imll}{\textsc{imll}}

\newcommand{\abs}{\lambda}
\newcommand{\app}{\mathbin{\cdot}}

\newcommand{\loc}{\ell}

\newcommand{\link}{\mathit{link}}

\newcommand{\zero}{\mathbf{0}}
\newcommand{\un}{I}
\newcommand{\paral}{\mid}

\newcommand{\transl}[1]{\llbracket #1 \rrbracket}

\newcommand{\id}{\mathrm{id}}
\newcommand{\iso}{\cong}
\newcommand{\impll}{\multimap}
\newcommand{\tens}{\otimes}
\newcommand{\bigtens}{\bigotimes}
\newcommand{\Bigtens}[1]{\displaystyle{\bigtens_{#1}}}

\newcommand{\card}[1]{|#1|}

\newcommand{\ens}[1]{\{ #1 \}}

\newcommand{\alt}{\mathrel{|}}
\newcommand{\name}[1]{\ulcorner #1 \urcorner}

\newcommand{\nop}[1]{#1^{\bot}}
\newcommand{\para}{\mathbin{|}}
\newcommand{\send}[2]{\ensuremath{\bar{#1}\langle#2\rangle}}
\newcommand{\get}[2]{\ensuremath{#1(#2)}}

\title{Variable binding, symmetric monoidal closed theories, and
  bigraphs}

\author{Richard Garner\inst{1} \and Tom Hirschowitz\inst{2} \and
  Aur\'elien Pardon\inst{3}} \institute{Cambridge University \and
  CNRS, Universit\'e de Savoie \and ENS Lyon}

\begin{document}
\maketitle
\begin{abstract}
  This paper investigates the use of symmetric monoidal closed
  (\smc{}) structure for representing syntax with variable binding, in
  particular for languages with linear aspects.  In this setting, one
  first specifies an \smc{} \emph{theory} $\theory$, which may express
  binding operations, in a way reminiscent from higher-order abstract
  syntax (\hoas{}). This theory generates an \smc{} category $S
  (\theory)$ whose morphisms are, in a sense, terms in the desired
  syntax. We apply our approach to Jensen and Milner's (abstract
  binding) bigraphs, in which \emph{processes} behave linearly, but
  \emph{names} do not. This leads to an alternative category of
  bigraphs, which we compare to the original.
\end{abstract}

\section{Introduction}
How to rigorously handle variable binding? The recent amount of
research on this issue attests its
delicacy~\cite{PittsAM:newaas,fiore:presheaf,hirscho:lam}. A main
difficulty is perhaps to reconcile $\alpha$-conversion with initial
algebra semantics: $\alpha$-conversion equates terms up to renaming of
bound variables; initial algebra semantics requires that terms form
the free, or initial model specified by a given signature.

We here investigate an approach sketched by Coccia et
al.~\cite{Coccia}, based on \smc{} theories, which they called
GS$\cdot \Lambda$ theories. In this setting, one first specifies an
\smc{} \emph{theory} $\theory$, which may express binding operations,
in a way reminiscent from higher-order abstract
syntax~\cite{PfenningElliott:hoas} (\hoas{}). This theory freely
generates an \smc{} category $S (\theory)$ whose morphisms are, in a
sense, terms in the desired syntax.

The known presentations of $S (\theory)$ mainly fall into two classes:
syntactic or graphical. Our emphasis in this paper is on a graphical
presentation of $S (\theory)$ and example applications.

We start in Section~\ref{sec:smctheories} with an expository account
of \smc{} theories and our construction of $S (\theory)$. This yields
a monadic adjunction. The morphisms of $S (\theory)$ are a variant of
proof nets, in the sense of intuitionistic multiplicative linear
logic~\cite{ll} (\imll{}): they are equivalence classes of special
graphs called \emph{linkings}, which must satisfy a certain
\emph{correctness} condition. Linkings compose by ``glueing'' the
graphs together, and correctness is preserved by composition.

We continue with a few examples in Section~\ref{sec:examples}, to
demonstrate the use of $S (\theory)$ as a representation for syntax
with variable binding.  In our presentation of $S (\theory)$, terms
look like the usual abstract syntax trees, and are actually a
generalisation of previous graphical forms of $\lambda$-calculus,
e.g., Wadsworth's $\lambda$-graphs~\cite{Wadsworth:phd}.  The objects
of $S (\theory)$ are \imll{} formulae. The subcategory of rank $0$
formulae (without $\impll$) roughly corresponds to terms, and
composition models linear substitution. Rank $1$ formulae express a
kind of term with holes, or \emph{context}, and composition models a
kind of constrained context application, or substitution with
capture. Formulae of higher ranks yield a form of higher-order
contexts. This kind of substitution would show up with any closed
structure, e.g., cartesian closed categories, but is not directly
available in more traditional
approaches~\cite{fiore:presheaf,hirscho:lam,PittsAM:newaas}.
Conversely, non-linear capture-avoiding substitution requires a bit
more work in our setting (and does not appear in this paper).  Along
the way, we prove a decomposition result showing the flexibility of
our approach, and we observe that the use of \smc{} structure
facilitates the cohabitation of linear and non-linear aspects in a
common language.

To further support this latter claim, Section~\ref{sec:big} studies
Jensen and Milner's bigraphs~\cite{Milner:bigraphs} in our setting, in
which \emph{processes} behave linearly, but \emph{names} do not.  We
translate each bigraphical signature $\bsig$ into an \smc{} theory
$\theory_\bsig$, and show that bigraphs over $\bsig$ essentially embed
into $S(\theory_\bsig)$, the free \smc{} category generated by
$\theory_\bsig$. Furthermore, although $S (\theory_\bsig)$ is much
richer than the original, the embedding is surjective on whole
programs.

\section{Symmetric monoidal closed theories}\label{sec:smctheories}
In this section, we provide an overview of the construction of $S
(\theory)$. A more technical presentation may be found in our
work~\cite{GHP}, which itself owes much to Trimble~\cite{Trimble:phd}
and Hughes~\cite{Hughes:freestar}.
\subsection{Signatures}
Roughly, an \smc{} category is a category with a tensor product
$\tens$ on objects and morphisms, symmetric in the sense that $A \tens
B$ and $B \tens A$ are isomorphic, and such that $(- \tens A)$ has a
right adjoint $(A \impll -)$, for each object $A$. We do not give
further details, since we are interested in describing the free such
category, which happens to be easier. Knowing that there is a category
$\SMCCat$ of \smc{} categories and strictly structure-preserving
functors should be enough to grasp the following.

An \smc{} \emph{signature} $\Sigma$ consists of a set $X$ of
\emph{sorts}, equipped with a (directed, multi-) graph whose vertices
are \imll{} formulae over $X$, as defined by:
$$\begin{array}{rcl@{\hspace*{2cm}}r}
  A, B, \ldots \in \Formulae (X) & ::= & x \alt \un \alt A \tens B \alt A \impll B & 
  \mbox{(where $x \in X$).}
\end{array}$$
We think of each edge $\to{A}{B}$ of the graph as specifying an operation 
of type $\to{A}{B}$.
A morphism of signatures $\to[1cm]{(X, \Sigma)}{(Y, \Sigma')}$ is a 
function $X \rTo^{f} Y$, equipped with a morphism of graphs, whose 
vertex component is ``$\Formulae (f)$'', i.e., the function sending 
any formula $A (x_1, \ldots, x_n)$ to $A (f (x_1), \ldots, f (x_n))$.
This defines a category $\Sig$ of signatures.

There is then a forgetful functor $\SMCCat \rTo^{U} \Sig$ sending each
\smc{} category $\C$ to the graph with as vertices formulae in
$\Formulae (\ob (\C))$, and as edges $\to{A}{B}$ the morphisms
$\to[.7cm]{\transl{A}}{\transl{B}}$ in $\C$, where $\transl{A}$ is
defined inductively to send each syntactic connective to the
corresponding function on $\ob (\C)$.

We will now construct an \smc{} category $S (\Sigma)$ from any
signature $\Sigma$, and extend this to a functor $\Sig \rTo^{S}
\SMCCat$, left adjoint to $U$.  How does $S (\Sigma)$ look like?
Under the Curry-Howard-Lambek correspondence, an \smc{} signature
amounts to a set of \imll{} axioms, and the free \smc{} category $S
(\Sigma)$ over a signature $\Sigma$ has as morphisms \imll{} proofs
under the corresponding axioms, modulo cut elimination. Or,
equivalently, morphisms are a variant of proof nets, which we
introduce gradually in the next sections.

\subsection{The free symmetric monoidal closed category over a set}
In the absence of axioms, i.e., given only a set of sorts, or
propositional variables, say $X$, Hughes~\cite{Hughes:freestar} has
devised a simple presentation of $S (X)$. Consider for a guiding
example the two endomorphisms of $((a \impll \un) \impll \un) \impll
\un$:
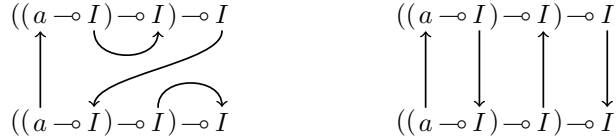
\begin{center}
  \begin{net}
    \matrix[column sep=0pt] (dom) { %
      \lpare{} & \lpare{} &\ssf{a}{a} & \gimpll{} & \ssf{i}{\un} &
      \rpare{} & %
      \gimpll{} & \ssf{ii}{\un} & \rpare{} & \gimpll{} &
      \ssf{iii}{\un}\\}; \matrix[column
    sep=0pt,at=(dom.south),below=1cm] (cod) { %
      \lpare{} & \lpare{} &\ssf{a'}{a} & \gimpll{} & \ssf{i'}{\un} &
      \rpare{} & %
      \gimpll{} & \ssf{ii'}{\un} & \rpare{} & \gimpll{} &
      \ssf{iii'}{\un}\\}; %
    \inetwire[->](i.south)(ii.south)
    \inetwire[->](iii.south)(i'.north)
    \inetwire[->](ii'.north)(iii'.north)
    \inetwire[->](a'.north)(a.south)
  \end{net}
\hfil
  \begin{net}
    \matrix[column sep=0pt] (dom) { %
      \lpare{} & \lpare{} &\ssf{a}{a} & \gimpll{} & \ssf{i}{\un} &
      \rpare{} & %
      \gimpll{} & \ssf{ii}{\un} & \rpare{} & \gimpll{} &
      \ssf{iii}{\un}\\}; %
    \matrix[column
    sep=0pt,at=(dom.south),below=1cm] (cod) { %
      \lpare{} & \lpare{} &\ssf{a'}{a} & \gimpll{} & \ssf{i'}{\un} &
      \rpare{} & %
      \gimpll{} & \ssf{ii'}{\un} & \rpare{} & \gimpll{} &
      \ssf{iii'}{\un} \\}; %
    \inetwire[->](i.south)(i'.north)
    \inetwire[->](ii'.north)(ii.south)
    \inetwire[->](iii.south)(iii'.north)
    \inetwire[->](a'.north)(a.south)
  \end{net}
\end{center}
(the right-hand one being the identity). 

First, the \emph{ports} of a formula, i.e., occurrences of sorts or of
$\un$, are given polarities: a port is \emph{positive} when it lies to
the left of an even number of $\impll$'s in the abstract syntax tree,
and \emph{negative} otherwise\footnote{The sign of a port in $A$ is
  directly apparent viewing $A$ is a classical LL formula, see the
  next paragraph.}.  For example, in the above formula, $a$ and the
middle $\un$ are negative, the other occurrences of $\un$ being
positive. When constructing morphisms $\to{A}{B}$, the ports in $A$
and $B$ will be assigned a \emph{global} polarity, or a polarity
\emph{in} the morphism: the ports of $B$ have their polarity in $B$,
while those of $A$ have the opposite polarity.  For example, in the
above examples, the occurrence of $a$ in the domain is globally
positive.

A \emph{linking} is a partial function $f$ from negative ports to
positive ports, such that for each sort $a$, $f$ maps negative $a$
ports to positive $a$ ports, bijectively. We observe that this allows
to connect $\un$ ports to ports of any type. This last bit does not
appear in the above example; it does in~\eqref{eq:rho} below.  Clearly
from the example, linkings are kind of graphs, and we call their edges
\emph{wires}.

A linking is then \emph{correct} when it is a total function, and when
it moreover satisfies the Danos-Regnier (\dr{})
criterion~\cite{Danos:mll}. The latter roughly goes as follows. An
\imll{} formula may be written using the connectives of
\emph{classical} linear logic, defined by the grammar:
$$\begin{array}{rcl@{\quad \mid \quad}l@{\quad \mid \quad}l}
A, B, \ldots & ::= & x    &  \un &  A \otimes B \\
&  \mid   & x^\bot & \bot & A \parr B.
\end{array}$$
The de Morgan dual $A^\bot$ of $A$ is defined as usual (by swapping 
connectives, vertically in the above grammar).
We have removed $A \impll B$, now encoded as $A^\bot \parr B$; some
classical formulae are not expressible in \imll{}, such as $\bot$, or
$x \parr x$. 
The classical formulation of our above example is $((\nop{a} \parr
\un) \tens \bot) \parr \un$.

Then, a \emph{switching} of a classical formula is its abstract syntax
tree, where exactly one argument edge of each $\parr$ has been
removed. A \emph{switching} of a linking $A \rTo^{f} B$ is a graph
obtained by glueing (in the sense of pushouts in the category of
undirected graphs) along ports the (undirected) wires of $f$ with
switchings of $\nop{A}$ and $B$. The linking then satisfies \dr{} iff
all its switchings are acyclic and connected.

On our above examples, a sample switching yields
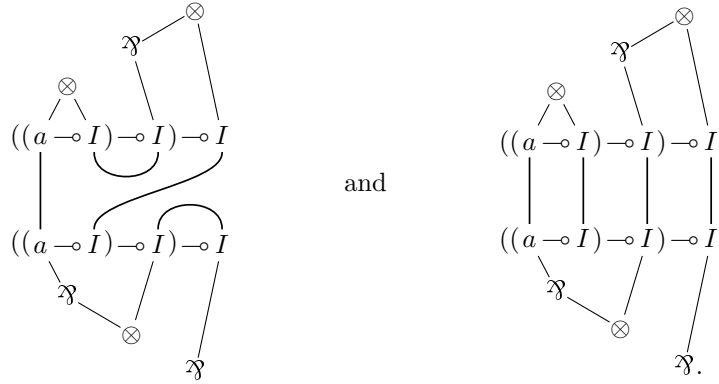
\begin{center}
  \begin{net}
    \matrix[column sep=0pt] (dom) { %
      \lpare{} & \lpare{} &\ssf{a}{a} & \gimpllof{parl} & \ssf{i}{\un} &
      \rpare{} & %
      \gimpllof{tens} & \ssf{ii}{\un} & \rpare{} & \gimpllof{parr} &
      \ssf{iii}{\un}\\}; \matrix[column
    sep=0pt,at=(dom.south),below=1cm] (cod) { %
      \lpare{} & \lpare{} &\ssf{a'}{a} & \gimpllof{parl'} & \ssf{i'}{\un} &
      \rpare{} & %
      \gimpllof{tens'} & \ssf{ii'}{\un} & \rpare{} & \gimpllof{parr'} &
      \ssf{iii'}{\un}\\}; %
    \inetwire(i.south)(ii.south)
    \inetwire(iii.south)(i'.north)
    \inetwire(ii'.north)(iii'.north)
    \inetwire(a'.north)(a.south)
    \node[at=(parl.center),above=.5cm] (parlu) {$\tens$} ;
    \node[at=(tens.center),above=1cm] (tensu) {$\parr$} ;
    \node[at=(parr.center),above=1.5cm] (parru) {$\tens$} ;
    \draw (parlu) -- (a) ;
    \draw (parlu) -- (i) ;
    \draw (tensu) -- (ii) ;
    \draw (parru) -- (iii) ;
    \draw (parru) -- (tensu) ;
    \node[at=(parl'.center),below=.5cm] (parlu') {$\parr$} ;
    \node[at=(tens'.center),below=1cm] (tensu') {$\tens$} ;
    \node[at=(parr'.center),below=1.5cm] (parru') {$\parr$} ;
    \draw (parlu') -- (a') ;
    \draw (tensu') -- (parlu') ;
    \draw (tensu') -- (ii') ;
    \draw (parru') -- (iii') ;
  \end{net}
\hfil and \hfil
  \begin{net}
    \matrix[column sep=0pt,at={(0,1.3)}] (dom) { %
      \lpare{} & \lpare{} &\ssf{a}{a} & \gimpllof{parl} & \ssf{i}{\un}
      & \rpare{} & %
      \gimpllof{tens} & \ssf{ii}{\un} & \rpare{} & \gimpllof{parr} &
      \ssf{iii}{\un}\\}; %
    \matrix[column sep=0pt,,at={(0,0)}] (cod) { %
      \lpare{} & \lpare{} &\ssf{a'}{a} & \gimpllof{parl'} &
      \ssf{i'}{\un} & \rpare{} & %
      \gimpllof{tens'} & \ssf{ii'}{\un} & \rpare{} & \gimpllof{parr'}
      & \ssf{iii'}{\un}\\}; %
    \inetwire(i.south)(i'.north) \inetwire(ii'.north)(ii.south)
    \inetwire(iii.south)(iii'.north) \inetwire(a'.north)(a.south)
    \node[at=(parl.center),above=.5cm] (parlu) {$\tens$} ;
    \node[at=(tens.center),above=1cm] (tensu) {$\parr$} ;
    \node[at=(parr.center),above=1.5cm] (parru) {$\tens$} ; \draw
    (parlu) -- (a) ; \draw (parlu) -- (i) ; \draw (tensu) -- (ii) ;
    \draw (parru) -- (iii) ; \draw (parru) -- (tensu) ;
    \node[at=(parl'.center),below=.5cm] (parlu') {$\parr$} ;
    \node[at=(tens'.center),below=1cm] (tensu') {$\tens$} ;
    \node[at=(parr'.center),below=1.5cm] (parru') {$\parr$} ; \draw
    (parlu') -- (a') ; \draw (tensu') -- (parlu') ; \draw (tensu') --
    (ii') ; %
    \draw (parru') -- (iii') ; %
    \point{parru'} %
  \end{net}
\end{center}

Linkings compose by glueing along ports in the middle formula, and
correctness is preserved under composition, which yields a category
$S_0 (X)$. For example, composing the structural isomorphism $\rho$
with its obvious candidate inverse yields:
\begin{equation}
  \begin{array}[c]{c@{\hspace*{1cm}}c@{\hspace*{1cm}}c}
  \begin{net}
    \matrix[column sep=0pt] (dom) { %
      \ssf{a}{a} & \gtens{} & \ssf{i}{\un} \\}; %
    \matrix[column sep=0pt,at=(dom.center),below=1cm] (cod) { %
      \ssf{a'}{a} &  & \node{\ \ \ \ \ \ }; \\}; %
    \matrix[column sep=0pt,at=(cod.center),below=1cm] (codd) { %
      \ssf{a''}{a} & \gtens{} & \ssf{i''}{\un} \\}; %
    \inetwire[->](i.south)(a'.north)
    \inetwire[->](a.south)(a'.north)
    \inetwire[->](a'.south)(a''.north)
  \end{net}& = & 
  \begin{net}
    \matrix[column sep=0pt] (dom) { %
      \ssf{a}{a} & \gtens{} & \ssf{i}{\un} & \node{$\ $}; \\}; %
    \matrix[column sep=0pt,at=(cod.center),below=1cm] (codd) { %
      \ssf{a''}{a} & \gtens{} & \ssf{i''}{\un} 
      \\}; %
    \inetwire[->](i.south)(a''.north) %
    \inetwire[->](a.south)(a''.north) %
    \point{i''} %
  \end{net}
\end{array}
\label{eq:rho}
\end{equation}

This does not yield an identity. And indeed, correct linkings do not
form an \smc{} category. Instead, they form the free \emph{split}
\smc{} category over $X$~\cite{Hughes:freestar}.  A split \smc{}
category is like an \smc{} category, where $\lambda$ and $\rho$ are
only required to have left inverses, as exemplified with $\rho$
in~\eqref{eq:rho}.

Here is the final step: let a \emph{rewiring} of some correct linking
$f$ be any linking obtained by changing the target of exactly one wire
from some occurrence of $\un$ in $f$, without breaking
correctness. Typically, \eqref{eq:rho} rewires to the identity.  Then
$S (X)$ is the result of quotienting $S_0 (X)$ by the equivalence
relation generated by rewiring.

\subsection{The free symmetric monoidal closed category over a
  signature}\label{subsec:free:sig}
We now extend the construction to \smc{} signatures $\Sigma$: we have
a set of sorts $X$, plus a set of operations.  We now enrich linkings
with, for each operation $A \rTo^{c} B$, a formal morphism. We picture
these by \emph{cells}, in the spirit of \emph{interaction
  nets}~\cite{Lafont90}.  For example, consider the $\pi$-calculus: we
will see in Section~\ref{sec:examples} that the corresponding
signature has one sort $v$ for names and one sort $t$ for processes,
and among others two operations send and get of types:
$$(v
\tens v \tens t) \rTo^{\pisend} t \hspace*{.2\linewidth} (v \tens (v
\impll t)) \rTo^{\piget} t.$$ This yields cells: \hfil
  \begin{net}
    \matrix[column sep=0pt] (dom) { %
      \ssf{v}{v} & \gtens{} & \ssf{v'}{v} & \gtens{} & \ssf{t}{t} \\}; %
    \matrix[column sep=0pt,at=(dom.center),below=2cm](cod){ \ssf{t'}{t}\\ }; %
    \inetcell[at=(dom.south),below=.5cm](send){$\pisend$} %
    \inetwire[->](t.south)(send.left pax)
    \inetwire[->](v'.south)(send.middle pax)
    \inetwire[->](v.south)(send.right pax)
    \inetwire[->](send.pal)(t'.north)
  \end{net}
\hfil and \hfil 
  \begin{net}
    \matrix[column sep=0pt] (dom) { %
      \ssf{v'}{v} & \gtens{} & \lpare{} & \ssf{v}{v} & \gimpll{} &
      \ssf{t}{t} & \rpare{} \\}; %
    \matrix[column sep=0pt,at=(dom.center),below=2cm](cod){
      \ssf{t'}{t} \\ }; %
    \inetcell[at=(dom.south),below=.5cm](send){$\piget$} %
    \inetwire[->](v'.south)(send.right pax)
    \inetwire[->](t.south)(send.left pax)
    \inetwire[<-](v.south)(send.middle pax)
    \inetwire[->](send.pal)(t'.north) %
    \point{t'} %
  \end{net}

\begin{figure}[t]
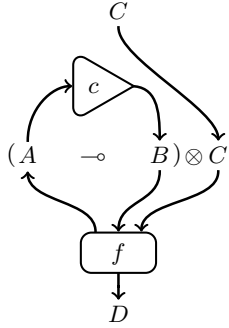

  \centering
  \begin{net}
    \matrix[column sep=0pt] (dom) { %
      & \ssf{C}{C} \\}; %
    \matrix[column sep=0pt,at=(dom.south),below=1.5cm] (mid) { %
      \lpare{} & \ssf{A}{A} &
      \ssf{impl}{\hspace*{.5cm}\impll\hspace*{.5cm}} & %
      \ssf{B}{B} & %
      \rpare{} & \gtens{} & \ssf{C'}{C} \\}; %
    \matrix[column sep=0pt,at=(mid.south),below=1.7cm] (cod) { %
      \ssf{D}{D} \\} ; %
    \inetcable[->](C.south)(C'.north) %
    \inetcell[at=(impl.north),above=.4cm](c){$c$}[90] %
    \inetcable[->](A.north)(c.pax) \inetcable[->](c.pal)(B.north)
    \inetmor[at=(mid.south),below=.8cm](f){f}; %
    \inetcable[->](f.right pax)(A.south) %
    \inetcable[->](B.south)(f.north) %
    \inetcable[->](C'.south)(f.left pax) %
    \inetcable[->](f.south)(D.north) %
  \end{net}
\caption{Functional completeness}
\label{fig:funcomp}
\end{figure}

  We then extend linkings $\to{A}{B}$ to include cells in a suitable
  way --- a glance at~\eqref{eq:glance} might help. We consider
  linking equivalent modulo the choice of support, i.e., the choice of
  cells.  Linkings compose as before. The question is then: what is a
  switching in the extended setting?  The answer is that taking a
  switching of a cell $A \rTo^{c} B$ is replacing the cell with a
  switching of $A \tens \nop{B}$. For example, consider the send and
  get operations, and a \emph{contraction} operation $v \rTo^{c} v
  \tens v$. Their respective switchings are:
  
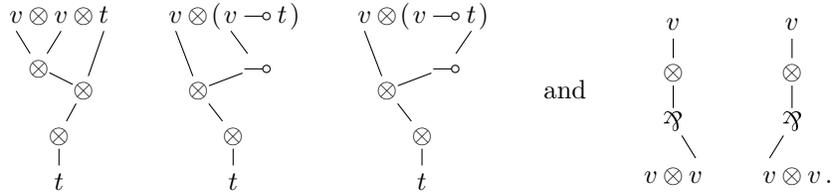
\begin{center}
  \begin{net}
    \matrix[column sep=0pt] (dom) { %
      \ssf{v}{v} & \gtensof{tensd} & \ssf{v'}{v} & \gtensof{tensd'} & %
      \ssf{t}{t} \\}; %
    \matrix[column sep=0pt,at=(dom.center),below=2cm](cod){ \ssf{t'}{t}\\ }; %
    \node[at=(tensd.south),below=.3cm] (tens) {$\tens$} ; %
    \node[at=(tensd'.south),below=.6cm] (tens') {$\tens$} ; %
    \node[at=(dom.south),below=1.2cm] (tens'') {$\tens$} ; %
    \draw (t.south) -- (tens') ; %
    \draw (tens) -- (tens') ; %
    \draw (v'.south) -- (tens) ; %
    \draw (v.south) -- (tens) ; %
    \draw (tens') -- (tens'') ; %
    \draw (tens'') -- (t'.north) ; %
  \end{net}
\hfil
  \begin{net}
    \matrix[column sep=0pt] (dom) { %
      \ssf{v'}{v} & \gtensof{tensd'} & \lpare{} & \ssf{v}{v} &
      \gimpllof{tensd} & \ssf{t}{t} & \rpare{} \\}; %
    \matrix[column sep=0pt,at=(dom.center),below=2cm](cod){
      \ssf{t'}{t}\\ }; %
    \node[at=(tensd.south),below=.3cm] (tens) {$\impll$} ; %
    \node[at=(tensd'.south),below=.6cm] (tens') {$\tens$} ; %
    \node[at=(dom.south),below=1.2cm] (tens'') {$\tens$} ; %
    \draw (tens) -- (tens') ; %
    \draw (v'.south) -- (tens') ; %
    \draw (v.south) -- (tens) ; %
    \draw (tens') -- (tens'') ; %
    \draw (tens'') -- (t'.north) ; %
  \end{net} \hfil
  \begin{net}
    \matrix[column sep=0pt] (dom) { %
      \ssf{v'}{v} & \gtensof{tensd'} & \lpare{} & \ssf{v}{v} &
      \gimpllof{tensd} & \ssf{t}{t} & \rpare{} \\}; %
    \matrix[column sep=0pt,at=(dom.center),below=2cm](cod){ \ssf{t'}{t}\\ }; %
    \node[at=(tensd.south),below=.3cm] (tens) {$\impll$} ; %
    \node[at=(tensd'.south),below=.6cm] (tens') {$\tens$} ; %
    \node[at=(dom.south),below=1.2cm] (tens'') {$\tens$} ; %
    \draw (t.south) -- (tens) ; %
    \draw (tens) -- (tens') ; %
    \draw (v'.south) -- (tens') ; %
    \draw (tens') -- (tens'') ; %
    \draw (tens'') -- (t'.north) ; %
  \end{net}
\hfil and \hfil
\begin{net}
    \matrix[column sep=0pt] (dom) { \ssf{v}{v} \\ } ; %
    \matrix[column sep=0pt,at=(dom.center),below=1.8cm](cod){ %
      \ssf{v'}{v} & \gtensof{tens} & \ssf{v''}{v} \\ }; %
    \node[at=(tens.north),above=.3cm] (parr) {$\parr$} ; %
    \node[at=(parr.north),above=.3cm] (tens') {$\tens$} ; %
    \draw (v.south) -- (tens') ; %
    \draw (tens') -- (parr) ; %
    \draw (v''.north) -- (parr) ; %
  \end{net}
\hfil
    \begin{net}
    \matrix[column sep=0pt] (dom) { \ssf{v}{v} \\ } ; %
    \matrix[column sep=0pt,at=(dom.center),below=1.8cm](cod){ %
      \ssf{v'}{v} & \gtensof{tens} & \ssf{v''}{v} \\ }; %
    \node[at=(tens.north),above=.3cm] (parr) {$\parr$} ; %
    \node[at=(parr.north),above=.3cm] (tens') {$\tens$} ; %
    \draw (v.south) -- (tens') ; %
    \draw (tens') -- (parr) ; %
    \draw (v'.north) -- (parr) ; %
    \point{v''} %
  \end{net}
\end{center}
To understand why this is right, observe that \smc{} categories have a
\emph{functional completeness} property, in the sense of Lambek and
Scott~\cite{Lambek:categorical}. Roughly, this means that any morphism
$\to{C}{D}$ using a cell $A \rTo^{c} B$ may be parameterised over it, i.e.,
be decomposed as
\begin{equation}
\begin{diagram}
C & \rTo^{\iso} & \un \tens C & &\rTo^{\name{c} \tens C} & & (A \impll B) \tens C
& \rTo^{f} & D,
\end{diagram}\label{eq:name}
\end{equation}
where $\name{c}$ is the currying of $c$.  Figure~\ref{fig:funcomp}
pictures~\eqref{eq:name} graphically. (Thick wires denote several
atomic wires in parallel.)  This rightly suggests that an operation $A
\rTo^{c} B$ should have the same switchings as $A \impll B$ in the
domain, i.e., $A \tens \nop{B}$.

\begin{theorem}
  This yields a monadic adjunction
  \begin{diagram}[width=1.4cm]
    \Sig & \pile{\rTo^{S} \\ \bot \\ \lTo_{U}} & \SMCCat.
  \end{diagram}
\end{theorem}

\subsection{The free symmetric monoidal closed category over a
  theory}\label{subsec:co:monoids}
That gives the construction for signatures. We now extend it to \smc\
theories: define a theory $\theory$ to be given by a signature $\sig$,
together with a set $E_{A,B}$ of equations between morphisms in $S
(\sig) (A, B)$, for all $A, B$.  The free \smc\ category $S (\theory)$
generated by such a theory is then the quotient of $S (\sig)$ by the
equations. Constructing $S (\theory)$ graphically is more direct than
could have been feared: we first define the binary predicate $f_1 \sim
f_2$ relating two morphisms
$C \pile{\rTo^{f_1, f_2} \\
  \rTo} D$ in $S (\sig)$ as soon as each $f_i$ decomposes
(remember~\eqref{eq:name} and Figure~\ref{fig:funcomp}) as
\begin{diagram}
  C & \rTo^{\iso} & \un \tens C && \rTo^{\name{g_i} \tens C} && (A \impll
  B) \tens C & \rTo^{f} & D
\end{diagram}
with a common $f$, with $(g_1, g_2) \in E_{A, B}$, and where
$\name{g}$ is the currying of $g$. Then, we take the smallest
generated equivalence relation, prove it stable under composition, and
quotient $S (\sig)$ accordingly.

Finally, $S (\theory)$ is initial in the following sense. Let the
category of \emph{representations} of $\theory$ be the full
subcategory of the comma category $\sig {\downarrow} U$ whose objects
are the morphisms $\to[.7cm]{\sig}{U(\C)}$, for which $\C$ is an
\smc{} category satisfying the equations in $E$. Now consider the
morphism
\begin{diagram}[inline,tight]
  \sig & \rTo^{\eta} & US (\sig) & \rTo^{q} & US (\theory),
\end{diagram}
where $q$ is the quotient by the equations in $E$.
\begin{theorem}
  This morphism is initial in the category of representations of
  $\theory$.
\end{theorem}

\subsection{Commutative monoid objects}
We finally slightly tune the above construction to better handle the
special case of commutative monoids. In a given theory $\theory =
(\sig, E)$, assume that a sort $t$ is equipped with two operations $t
\tens t \rTo^{m} t$ and $\un \rTo^{e} t$, with equations making it
into a commutative monoid ($m$ is associative and commutative, $e$ is
its unit).  Further assume that $m$ and $e$ do not occur in other
equations. In this case, we sketch (for lack of space) an alternative,
more economic description of morphisms in $S (\theory)$.

Start from the original definition, relax the bijection condition on
linkings, i.e., allow them to map negative $a$ ports to positive $a$
ports non-bijectively for any $a$, and then replace $m$ and $e$ as
follows:
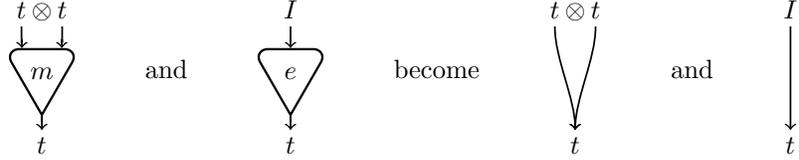
\begin{center}
  \begin{net}
    \inetcell[at={(0,1)}](m){$m$} %
    \matrix[column sep=0pt,at={(0,1.8)}] (dom) { %
      \ssf{t}{t} & \gtens{} & \ssf{t'}{t} \\}; %
    \matrix[column sep=0pt,at={(0,0)}] (cod) { %
      \ssf{t''}{t}\\ }; %
    \inetwire[->](t'.south)(m.left pax) \inetwire[->](t.south)(m.right
    pax) \inetwire[->](m.pal)(t''.north)
  \end{net}  \hfil and \hfil 
  \begin{net}
    \matrix[column sep=0pt,at={(0,1.8)}] (dom) { %
      \gun{i} \\}; %
    \matrix[column sep=0pt,at={(0,0)}](cod){ \ssf{t}{t}\\ }; %
    \inetcell[at={(0,1)}](e){$e$} %
    \inetwire[->](i.south)(e.middle pax)
    \inetwire[->](e.pal)(t.north)
  \end{net}  
  \hfil become \hfil
  \begin{net}
    \matrix[column sep=0pt,at={(0,1.8)}] (dom) { %
      \ssf{t}{t} & \gtens{} & \ssf{t'}{t} \\}; %
    \matrix[column sep=0pt,at={(0,0)}](cod){ \ssf{t''}{t}\\ }; %
    \inetwire[->](t.south)(t''.north)
    \inetwire[->](t'.south)(t''.north)
  \end{net}  
  \hfil  and \hfil 
    \begin{net}
    \matrix[column sep=0pt,at={(0,1.8)}] (dom) { %
      \gun{i} \\}; %
    \matrix[column sep=0pt,at={(0,0)}](cod){ \ssf{t}{t}\\ }; %
    \inetwire[->](i.south)(t.north) %
    \point{t} %
  \end{net}  
\end{center}

For a commutative comonoid $(c, w)$, the dual trick does not quite
work, because of problems with weakening. But still, a non-empty tree
of $c$'s may be represented by several arrows leaving its
root. Observe that while $m$ has as only switching the complete graph,
$c$ has two switchings (the formula is $v \tens (\nop{v} \parr
\nop{v})$).

\subsection{Modularity}
Melli\`es~\cite{Mellies} convincingly explains the need for
\emph{modular} models of programming languages and calculi.  In a
slightly different sense, we argue that \smc{} categories provide a
modular model of syntax. Namely, we obtain, for any theory $\theory$:
\begin{proposition}\label{prop:mod}
  For any proof net $A \rTo^{f} B$ in $S (\theory)$ with a set $C$ of
  cells, and any partition of $C$ into $C_1$ and $C_2$, $f$
  decomposes as
  \begin{diagram}
    A & \rTo^{f_1} & D & \rTo^{f_{2}} & B,
  \end{diagram}
  where each $f_i$ contains exactly the cells in $C_i$.
\end{proposition}
The proof is by inductively applying the decomposition in
Figure~\ref{fig:funcomp}. The proposition intuitively says that,
thinking of operations in $\sig$ as atomic building blocks, each term
may be obtained by plugging such blocks together by composition.  An
example is in Section~\ref{subsec:ex:mod}.

\section{First examples}\label{sec:examples}
In this section, we explain how to build the $\lambda$-calculus in
stages, starting from the linear $\lambda$-calculus, and passing
through a kind of $\lambda$-calculus with sharing of terms. We then
give an example application of Proposition~\ref{prop:mod}.  We end
with a $\pi$-calculus example, which we will use as our main example
in Section~\ref{sec:big}. 

\subsection{Lambda-calculus}
We start with the easiest application: the untyped $\lambda$-calculus.
If we naively mimick \hoas{} to guess a signature for the
$\lambda$-calculus, we obtain one sort $t$ and operations $t \tens t
\rTo^{\app} t$ and $(t \impll t) \rTo^{\abs} t$.  However, the free
\smc{} category on this signature is the \emph{linear}
$\lambda$-calculus, as shown by the following standard result:
\begin{proposition}
  Morphisms $\to{\un}{t}$ are in bijection with 
  closed linear $\lambda$-terms.
\end{proposition}
Composition in our category is like \emph{context} application in
$\lambda$-calculus. A context is a term with (possibly several,
numbered) holes, and context application is replacement of the hole
with a term (or another context), possibly capturing some
variables. The correspondence is tedious to formalise though, because
contexts do not have enough information. For example, consider the
context $\lambda x. (\trou_0 \app \trou_1)$ with two holes $\trou_0$
and $\trou_1$. Exactly one of $\trou_0$ and $\trou_1$ may use $x$, but
this information is not contained in the context, which makes context
application partial. In our setting, each possibility corresponds to
one of the following morphisms:
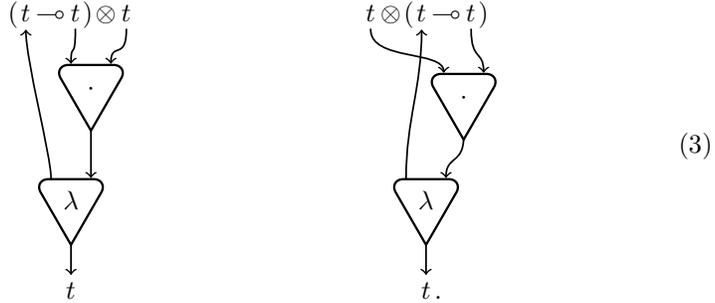
\begin{equation}
  \begin{net}
    \matrix[at={(0,4)}]{\lpare{} & \ssf{t}{t} & \gimpll{} &
      \ssf{t'}{t} & \rpare{} & \gtens{} & \ssf{t''}{t} \\ };%
    \matrix[at={(0,.3)}]{\ssf{t'''}{t} \\ };%
    \inetcell[at={(0,1.5)}](abs){$\abs$} %
    \inetcell[at={(abs.left pax)},above=.6cm](app'){$\app$} %
    \inetwire[->](abs.right pax)(t.south) %
    \inetwire[->](t'.south)(app'.right pax) %
    \inetwire[->](t''.south)(app'.left pax) %
    \inetwire[->](app'.pal)(abs.left pax) %
    \inetwire[->](abs.pal)(t'''.north) %
  \end{net}
\hspace*{.25\linewidth}
  \begin{net}
    \matrix[at={(0,4)}]{ \ssf{t'}{t} & \gtens{} & \lpare{} & \ssf{t}{t}
      & \gimpll{} & \ssf{t''}{t}  & \rpare{}\\ };%
    \matrix[at={(0,.3)}]{\ssf{t'''}{t} \\ };%
    \inetcell[at={(0,1.5)}](abs){$\abs$} %
    \inetcell[at={(.5,2.9)}](app'){$\app$} %
    \inetwireext[->](abs.right pax) to[out=90,in=-90] (t.south) ;%
    \inetwire[->](t'.south)(app'.right pax) %
    \inetwire[->](t''.south)(app'.left pax) %
    \inetwire[->](app'.pal)(abs.left pax) %
    \inetwire[->](abs.pal)(t'''.north) %
    \point{t'''}
  \end{net}\label{eq:glance}
\end{equation}

A first attempt to recover the full $\lambda$-calculus is to add a
contraction and a weakening $t \rTo^{c} t \tens t$ and $t \rTo^{w}
\un$ to our signature, with the equations making $(c, w)$ into a
commutative comonoid. The free \smc{} category on this theory
is close to Wadsworth's $\lambda$-graphs~\cite{Wadsworth:phd}, which
are a kind of $\lambda$-terms with a fine representation of
sharing. For example, it contains two morphisms
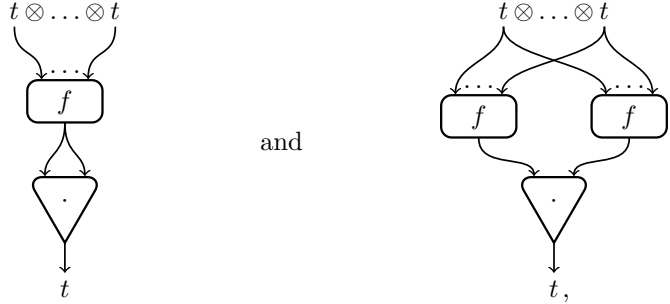
\begin{center}
  \begin{net}
    \matrix[at={(0,4)}]{
      \ssf{t}{t} & \ssf{}{\tens \ldots \tens} & \ssf{t'}{t} \\} ;%
    \matrix[at={(0,.3)}]{ 
      \ssf{t''}{t} \\} ;%
    \inetmor[at={(0,2.8)}](f){f}; %
    \inetcell[at={(0,1.5)}](app){$\app$} %
      \inetwire[->](f.south)(app.right pax) %
      \inetwire[->](f.south)(app.left pax) %
      \inetwire[->](app.pal)(t''.north) %
      \inetwire[->](t.south)(f.right pax) %
      \inetwire[->](t'.south)(f.left pax) %
      \node[at={(f.north)},anchor=south] {$\ldots$} ; %
  \end{net}
\hfil and \hfil 
  \begin{net}
    \matrix[at={(0,4)}]{
      \ssf{t}{t} & \ssf{}{\tens \ldots \tens} & \ssf{t'}{t} \\} ;%
    \matrix[at={(0,.3)}]{ 
      \ssf{t''}{t} \\} ;%
    \inetmor[at={(-1,2.6)}](f){f}; %
    \inetmor[at={(1,2.6)}](f'){f}; %
    \inetcell[at={(0,1.5)}](app){$\app$} %
      \inetwire[->](f.south)(app.right pax) %
      \inetwire[->](f'.south)(app.left pax) %
      \inetwire[->](app.pal)(t''.north) %
      \inetwire[->](t.south)(f.right pax) %
      \inetwire[->](t'.south)(f.left pax) %
      \node[at={(f.north)},anchor=south] {$\ldots$} ; %
      \inetwire[->](t.south)(f'.right pax) %
      \inetwire[->](t'.south)(f'.left pax) %
      \node[at={(f'.north)},anchor=south] {$\ldots$} ; %
      \virgule{t''}%
  \end{net}
\end{center}
which, because contraction is not natural, are different. 

To recover the standard $\lambda$-calculus without sharing, a solution
is to consider two sorts: a sort $t$ for terms, and a sort $v$ for
variables, an idea that has been explored in the context of
\hoas{}~\cite{dedou}. The theory then contains:
\begin{center}
  $t \tens t \rTo^{\app} t$ \hfil
  $(v \impll t) \rTo^{\abs} t$ \hfil
  $v \rTo^{c} v \tens v$ \hfil $v \rTo^{w} \un$ \hfil $v \rTo^{d} t$,
\end{center}
where the latter is instantiation of a variable as a term, and $(c,
w)$ is a commutative comonoid.  We obtain:
\begin{proposition}
  Morphisms $\to{\un}{t}$ are in bijection with closed
  $\lambda$-terms. Among them, those not using $c$ nor $w$ are in
  bijection with closed linear $\lambda$-terms.
\end{proposition}

\subsection{Higher order and modularity}\label{subsec:ex:mod}
We now give an example decomposition as in
Proposition~\ref{prop:mod}. Consider the context with numbered holes
$(\trou_0 \app \trou_1) \app \trou_3$, and consider the decomposition
of Proposition~\ref{prop:mod} with $C_1$ containing exactly the
outermost application. Pictorially,
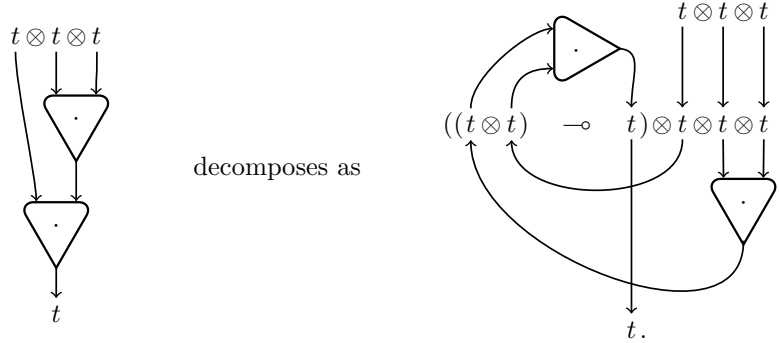
\begin{center}
  \begin{net}
    \matrix[at={(0,4)}]{\ssf{t}{t} & \gtens{} & \ssf{t'}{t} & \gtens{}
      & \ssf{t''}{t} \\ };%
    \matrix[at={(0,.3)}]{\ssf{t'''}{t} \\ };%
    \inetcell[at={(0,1.5)}](app){$\app$} %
    \inetcell[at={(app.left pax)},above=.5cm](app'){$\app$} %
    \inetwire[->](t.south)(app.right pax) %
    \inetwire[->](t'.south)(app'.right pax) %
    \inetwire[->](t''.south)(app'.left pax) %
    \inetwire[->](app'.pal)(app.left pax) %
    \inetwire[->](app.pal)(t'''.north) %
  \end{net} \hfil decomposes as \hfil
  \begin{net}
    \matrix[anchor=east,at={(0,4.5)}]{\ssf{t}{t} & \gtens{} &
      \ssf{t'}{t} & \gtens{} & \ssf{t''}{t} \\ };%
    \matrix[column sep=0pt,at={(0,3)},anchor=east] (mid) { %
      \lpare{} & \lpare{} & \ssf{tmid}{t} & \gtens{} & \ssf{tmid'}{t}
      & \rpare{} & \ssf{impl}{\hspace*{.4cm}\impll\hspace*{.4cm}} & %
      \ssf{tmid''}{t} & %
      \rpare{} & \gtens{} & \ssf{tm}{t} & \gtens{} & \ssf{tm'}{t} &
      \gtensof{tens} & \ssf{tm''}{t} \\}; %
    \matrix[column sep=0pt,at=(tmid''.south),below=2.3cm] (cod) { %
      \ssf{tcod}{t} \\} ; %
    \inetwire[->](t.south)(tm.north) %
    \inetwire[->](t'.south)(tm'.north) %
    \inetwire[->](t''.south)(tm''.north) %
    \inetwire[->](tmid''.south)(tcod.north) %
    \inetcell[at={(-2.6,4)}](c){$\app$}[90] %
    \inetcell[at=(tens.south),below=.5cm](app'){$\app$} %
    \inetwire[->](tmid.north)(c.left pax) %
    \inetwire[->](tmid'.north)(c.right pax) %
    \inetwire[->](c.pal)(tmid''.north) %
    \inetwireext[->](tm''.south) to[out=-90,in=90] (app'.left pax) ;%
    \inetwireext[->](tm'.south) to[out=-90,in=90] (app'.right pax) ;%
    \inetwireext[->](tm.south) to[out=-90,in=-90] (tmid'.south) ;%
    \inetwireext[->](app'.pal) to[out=-90,in=-90] (tmid.south) ;%
    \point{tcod} %
  \end{net}
\end{center}
Observe that this makes use of a higher-order formula, namely $(t
\tens t) \impll t$. Also observe in advance that Jensen and Milner's
category of bigraphs $\bbig (\bsig)$ does not feature such a
decomposition.

\subsection{Pi-calculus example}\label{subsec:net:ex}
A reasonable theory $\theory$ for the $\pi$-calculus could have at
least the operations $\pisend{}$ and $\piget{}$ specified above, plus
commutative comonoid structure $(c, w)$ on $v$, plus commutative
monoid structure $({\paral}, \zero)$ on $t$.  Consider furthermore a
name restriction operation $\un \rTo^{\nu} v$, with the equation $w
\circ \nu = \id_{\un}$. We do not claim that this theory $\theory$ is
the right one for the $\pi$-calculus, but it is relevant for bigraphs.
(An alternative type for $\nu$ is $\to[.8cm]{(v \impll t)}{t}$.)

\begin{figure}[t]
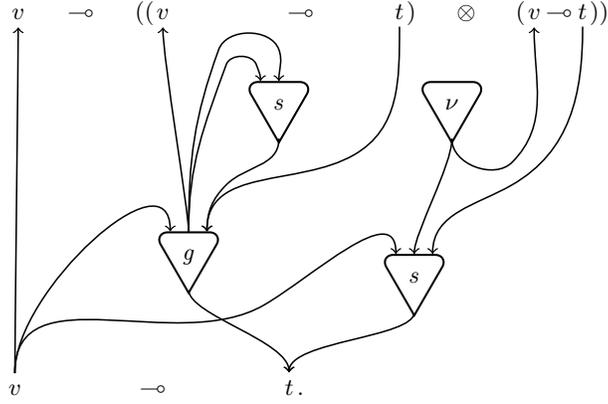

  \centering
  \begin{net}
    \matrix[column sep=0pt,at={(.6,4.5)}] (dom) { %
      \ssf{v}{v} &
      \ssf{impl}{\hspace*{0.5cm}\impll\hspace*{0.5cm}} %
      & \lpare{} & \lpare{} & %
      \ssf{v'}{v} &
      \ssf{impldom}{\hspace*{1.5cm}\impll\hspace*{1cm}}%
      & \ssf{t}{t} & \rpare{} & %
      \ssf{tensdom}{\hspace*{.5cm}\tens\hspace*{.5cm}} %
      & \lpare{} & \ssf{v''}{v} & \gimpll{} & \ssf{t'}{t} & \rpare{}
      & \rpare{}\\}; %
    \matrix[column sep=0pt,at={(-1.5,-.5)}] (cod) { %
      \ssf{v'''}{v} &
      \ssf{impl}{\hspace*{1.5cm}\impll\hspace*{1.5cm}} %
      & \ssf{t''}{t} \\}; %
    \inetcell[at={(2,1)}](snd'){$\pisend$} %
    \inetcell[at={(-1,1.3)}](get){$\piget$} %
    \inetcell[at={(.2,3.3)}](snd){$\pisend$} %
    \inetcell[at={(2.5,3.3)}](nu){$\nu$} %
    \inetwire[->](v'''.north)(v.south)%
    \inetwireext[->](v'''.north) to[out=90,in=90] (get.right pax) ;%
    \inetwireext[->] (v'''.north) to[out=90,in=210] (0,.6)
    to[out=30,in=90] (snd'.right pax) ; %
    \inetwire[->](get.middle pax)(v'.south) %
    \inetwireext[->] (get.middle pax) to[out=90,in=-110] (-.5,3.8)
    to[out=70,in=110] (-.1,3.8) %
    to[out=-70,in=90] (snd.right pax) ; %
    \inetwireext[->] (get.middle pax) to[out=90,in=-110] (-.6,4)
    to[out=70,in=110] (.2,4) %
    to[out=-70,in=90] (snd.middle pax) ; %
    \inetwireext[->] (snd.pal) to[out=-90,in=45] (-.5,2.2)
    to[out=-135,in=90] (get.left pax); %
    \inetwireext[->] (t.south) to[out=-90,in=30] (1.2,2.6)
    to[out=210,in=90] (get.left pax) ; %
    \inetwire[->](snd'.pal)(t''.north) %
    \inetwireext[->](nu.pal) to[out=-90,in=-135] (3.3,2.5)
    to[out=45,in=-90] (v''.south) ;%
    \inetwire[->](nu.pal)(snd'.middle pax) %
    \inetwireext[->] (t'.south) to[out=-90,in=30] (3.3,2.2)
    to[out=210,in=90] (snd'.left pax) ; %
    \inetwire[->](get.pal)(t''.north) %
    \point{t''} %
  \end{net}
  \caption{A $\pi$-calculus example}
\label{fig:main:ex}
\end{figure}
Consider the $\pi$-calculus term with ordered holes
$$(\get{a}{x}. (\trou_0 \para \send{x}{x})) \para %
\nu b. (\send{a}{b}.\trou_1).$$ This term may have many different
interpretations as a morphism in $S (\theory)$.  A first possibility
is depicted in Figure~\ref{fig:main:ex}.  Recall that several arrows
leaving a port mean a tree of contractions (the port has to have type
$v$), while several arrows entering a port mean a tree of parallel
compositions (the port has type $t$).  Finally, a positive $t$ port
with no entering arrow means a $0$.

The free variable $a$ of the term is represented by the occurrence of
$v$ in the codomain. It is used three times: twice following the term,
and once more for transmitting it to $\trou_0$ and $\trou_1$. 

But the language of \smc{} categories allows additional flexibility
w.r.t.\ syntax. For example, we could choose to impose that $\trou_0$
and $\trou_1$ may not use $a$. That would mean changing the domain for
$(v \impll t) \tens (v \impll t)$, and removing the leftmost wire.
Or, we could, e.g., only allow $\trou_0$ to use $a$, and not
$\trou_1$. That would only mean change the domain to $((v \tens v)
\impll t) \tens (v \impll t)$ (the leaves do not change, so the wires
may remain the same).

\section{Binding bigraphs}\label{sec:big}

In this section, we consider Jensen and
Milner's~\cite{Milner:bigraphs} (abstract binding) bigraphs.  They are
a general framework for reasoning about distributed and concurrent
programming languages, designed to encompass both the
$\pi$-calculus~\cite{Milner:pi} and the Ambient
calculus~\cite{Ambients}.  We are here only concerned with bigraphical
syntax: given what we call a \emph{bigraphical signature} $\bsig$,
Milner constructs a \emph{pre-category}, and then a category $\bbig
(\bsig)$, whose objects are \emph{bigraphical interfaces}, and whose
morphisms are bigraphs.

Its main features are (1) the presence of \emph{relative pushouts}
(RPOs) in the pre-category, which makes it well-behaved w.r.t.\
bisimulations, and that (2) in both the pre-category and the category,
the so-called \emph{structural} equations become equalities. Examples
of the latter are, e.g., in $\pi$ and Ambients, renaming of bound
variables, associativity and commutativity of parallel composition, or
scope extrusion for restricted names. Also, bigraphs follow a scoping
discipline ensuring that, roughly, bound variables are only used below
their binder.

We now proceed to recall what bigraphs are, and sketch our
interpretation in terms of \smc{} theories.

\subsection{Bigraphs}
We work with a slightly twisted definition of bigraphs, in two
respects.  First, we restrict Jensen and Milner's \emph{scope} rule by
adding a \emph{binding} rule to be respected by bigraphs. This rule
rectifies a deficiency of the scope rule, which prevented bigraphs to
be stable under composition in the original
paper~\cite{Milner:bigraphs}. It was added in later
work~\cite{Milner:bigraphs2}. Our second twist is to take names in an
infinite and totally ordered set of names fixed in advance, say
$\X$. This helps comparing bigraphs with our \smc{} category.

A \emph{bigraphical signature} is a set of operations, or
\emph{controls} $k \in \bsig$, with arity given by a pair of natural
numbers $a_k = (B_{k}, F_{k}) = (n, m)$. The number $B_k = n$ is the
number of \emph{binding} ports of $k$, $F_k = m$ being its number of
\emph{free} ports.  Additionally, a signature specifies a set $\A
\subseteq \bsig$ of \emph{atomic} controls, whose binding arity has to
be $0$.

Typically, get and send have arities: $a_{\pisend} = (0,2)$ and
$a_{\piget} = (1,1)$. They are not atomic (send would be atomic in the
asynchronous $\pi$-calculus). The other operations of the
$\pi$-calculus are all kind of built into bigraphical structure, as we
will see shortly.

Bigraphs form a category, whose objects are \emph{interfaces}.  An
interface is a triple $U = (n, X, \loc)$, where $n$ is a natural
number, $X \subseteq \X$ is a finite set of names, and $X \rTo^{\loc} n
+ \ens{\bot}$ is a \emph{locality} map ($n$ is identified with the set
$\ens{0, \ldots, n-1}$, i.e., the ordinal $n$). Names $x$ with $\loc
(x) = i \in n$ are \emph{located} at $i$; others are \emph{global}.

Introducing the morphisms, i.e., bigraphs, themselves seems easier by
example. We thus continue with an example bigraph, which will
correspond to the proof net in Figure~\ref{fig:main:ex}:
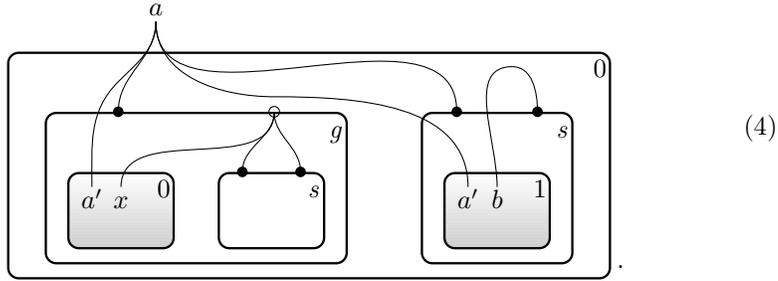
  \begin{equation}
    \begin{net}
      \bigraphplace[at={(2.5,0.6)}](whole){\ }[3cm,8cm] %
      \deco{whole}{0} %
      \bigraphplace[at={(1,0.3)}](get){\ }[2cm,4cm] %
      \deco{get}{\piget} %
      \bigraphplace[at={(2,0)}](snd){\ } %
      \deco{snd}{\pisend} %
      \bigraphplace[hole,at={(0,0)}](p1){\ } %
      \deco{p1}{0} %
      \bigraphplace[at={(5,0.3)}](snd'){\ }[2cm,2cm] %
      \deco{snd'}{\pisend} %
      \bigraphplace[hole,at={(5,0)}](p2){\ } %
      \deco{p2}{1} %
      \node[coordinate] (b) at ($(snd'.north) + (.2,.6cm)$) {}; %
      \draw (snd'.left pax) to[out=90,in=0] (b) ; %
      \fill (snd'.left pax) circle (2pt) ; %
      \draw ($(p2.north) - (0,.2cm)$) node[anchor=north] {$b$}
      to[out=90,in=180] (b) ; %
      \node[anchor=south] (a) at ($(whole.right pax) + (0,.4cm)$)
      {$a$} ; %
      \draw (a) to[out=-90,in=90] (snd'.right pax) ; %
      \fill (snd'.right pax) circle (2pt) ; %
      \draw (a) to[out=-90,in=90] (get.right pax) ; %
      \fill (get.right pax) circle (2pt) ; %
      \draw (a) to[out=-90,in=45] ($(get.right pax) + (0,.4cm)$)
      to[out=-135,in=90] node[anchor=north,pos=1] {$a'$} ($(p1.right
      pax) - (0,.2cm)$) ; %
      \draw (get.left pax) to[out=-90,in=90] (snd.left pax) ; %
      \fill (snd.left pax) circle (2pt) ; %
      \draw (get.left pax) to[out=-90,in=90] (snd.right pax) ; %
      \fill (snd.right pax) circle (2pt) ; %
      \draw (get.left pax) circle (2pt) ; %
      \draw (get.left pax) to[out=-90,in=90] node[anchor=north,pos=1]
      {$x$} ($(p1.north) - (0,.2cm)$) ; %
      \draw (a) to[out=-90,in=180] ($(get.left pax) + (0,.2cm)$)
      to[out=0,in=90] node[anchor=north,pos=1] {$a'$} ($(p2.right pax)
      - (0,.2cm)$) ; %
      \node[at=(whole.south east),anchor=south west,inner sep=2pt] {.}
      ; %
    \end{net}\label{eq:ex:big}
  \end{equation}
The codomain of this bigraph, which is graphically its outer face, is
$W = (1, \ens{a}, \ens{a \mapsto \bot})$: the element $0 \in 1$
represent the (only) outer box, which we accordingly marked $0$.  The
global name $a$ is the common end of the group of four wires reaching
the exterior of the box.

The domain of our example bigraph, which is graphically its inner face
when the grey parts are thought of as holes, is $U = (2, \ens{a', x,
  b}, \ens{x \mapsto 0, b \mapsto 1, a' \mapsto \bot})$. Comparing
this to the domain of our morphism in Figure~\ref{fig:main:ex}, we
observe that the elements $0$ and $1$ of $2$ correspond to $\trou_0$
and $\trou_1$.  Furthermore, the name $a'$ being global corresponds to
the domain $v \impll ((v \impll t) \tens (v \impll t))$
of Figure~\ref{fig:main:ex} having both $t$'s \emph{under the scope} of the
first $v$ (i.e., there is a $\impll$ with $t$ on its right, $v$ on its
left, and no other implication on the paths from it to them). Finally,
the locality map sending $x$ to $0$ corresponds to the second $v$
having only the first $t$ under its scope, and similarly for $b$ being
sent to $1$.

The morphism itself is a compound of two graphical structures. The
first structure, the \emph{place} graph, is a forest (here a tree),
whose leaves are the inner $0$ and $1$, the \emph{sites}, and whose
root is the outer $0$. Following Milner and Jensen, we represent nodes
by regions in the plane, the parent of a region being the immediately
enclosing region. The second structure, the \emph{link} graph, is a
bit more complicated to formalise. First, each internal (i.e., non
leaf, non root) node $v$ is labelled with an operation $k_v \in
\bsig$. We then compute the set of \emph{ports} $P$: it is the set of
pairs $(v, i)$, where $v$ is a node, and $i \in B_{k_v} + F_{k_v}$ is
in either component of the arity of $v$. The link graph is then a
function $P + X \rTo^{\link} E + Y,$ where $X = \ens{a', x, b}$ is the
set of \emph{inner} names, $Y = \ens{a}$ is the set of \emph{outer}
names, and $E$ is the set of \emph{edges}. In our morphism, $a'$ and
both occurrences of $a$ are mapped to the outer name $a$ by the
$\link$ map. Furthermore, $E$ is a two-element set, say $\ens{x',
  b'}$.  The edge $x'$ links the name $x$ received by the get node $g$
to its three occurrences. Formally, the three involved ports and the
name $x$ are all sent to $x'$ by the $\link$ map. The edge $b'$
represents the $\nu b$ in the term; formally, both $b$ and the
involved port of the right-hand $s$ node are sent to $b'$ by the
$\link$ map.

Until now, there is not much difference between the edge representing
the bound name $x$ received on $a$ and the bound name $b$ created by
$\nu b$.  The difference comes in when we check the \emph{scope} and
\emph{binding} rules. The binding rule requires that each binding port
(such as the one marked with a circle in~\eqref{eq:ex:big}) be sent to
an edge, as opposed to a name in the codomain. The scope rule further
requires that its \emph{peers}, i.e., the ports and names connected to
the same edge, lie strictly below it in the place graph. For ports,
this should be clear.  For inner names, this means that they should be
located at some site below it. In our example, the inner $0$ node
indeed lies below the get node, for instance. This all ensures that
bound names are only used below their binder.
\begin{remark}\label{rem:unique:binder}
  An edge is connected to at most one binding port, by acyclicity of
  the place graph. An edge connected to one binding port is called
  \emph{bound}.
\end{remark}

Composition $g \circ f$ in the category of bigraphs $\bbig (\bsig)$ is
by plugging the outer boxes of $f$ into the inner boxes of $g$, in
order, and connecting names straightforwardly. This only works if we
quotient out bigraphs by the natural notion of isomorphism, i.e.,
modulo choice of nodes and edges. We actually consider a further
quotient: removing an edge from $E$ which was outside the image of
$\link$.  The whole is called \emph{lean support} equivalence by
Jensen and Milner.

\subsection{Bigraphs as \smc{} theories}\label{subsec:trans}
We now describe our \smc{} theory for bigraphs, starting with the
translation of signatures.  Consider any signature
$(\bsig,B,F,\A)$. We translate it into the following \smc{} signature
$\theory_\bsig$, which has two sorts $\{t,v\}$, standing for terms and
variables (or names), and whose operations consist of
\emph{structural} operations and equations, plus \emph{logical}
operations.  The \emph{structural} part, accounting for the built-in
structure of bigraphs, is as in Section~\ref{subsec:net:ex}, i.e., it
consists of
\begin{itemize}
\item a commutative monoid structure $(\paral, \zero)$ on $t$,
\item a commutative comonoid structure $(c, w)$ on $v$, and
\item a name restriction $\un \rTo^{\nu} v$, such that $w \circ \nu =
  \id_{\un}$.
\end{itemize}
The \emph{logical} part consists, for each $k \in \bsig$ with $a_k =
(n, m)$, of an operation
  \begin{diagram}[inline,width=.9cm,midshaft]
    v^{\tens m} & \rTo^{k} & t
  \end{diagram} if $k$ is atomic (in which case $n = 0$), and
  \begin{diagram}[inline,width=4em,midshaft]
    (v^{\tens n} \impll t) \tens v^{\tens m} & \rTo^{k} & t
  \end{diagram} otherwise.

  For example, recall send and get, defined above to have arities
  $(1,1)$ and $(0,2)$, this gives exactly the operations $(v \tens v
  \tens t) \rTo^{\pisend} t$ and $(v \tens (v \impll t)) \rTo^{\piget}
  t$ from Section~\ref{subsec:free:sig}. An atomic get operation, in
  the style of the asynchronous $\pi$-calculus, would have the same
  bigraphical arity, translated into $v \tens v \rTo^{g'} t$.

Now, on objects, we define our functor $\T$ by:
\begin{equation}
  \T ( n,X,\loc)  = v^{\tens n_g} \impll \Bigtens{i \in n} ( v^{\tens n_i} \impll t),
\end{equation}\label{eq:T:ob}
where $n_g = \card{\loc^{-1}(\bot)}$ and for all $i \in n$, $n_i =
\card{\loc^{-1}(i)}$.  The ordering on $\X$ induces a bijection
between $X$ and $v$ leaves in the formula, which the translation of
morphisms exploits.  On our main example, this indeed maps the domain
and codomain of~\eqref{eq:ex:big} to those of Figure~\ref{fig:main:ex}.

We will here only describe the translation of morphisms
on~\eqref{eq:ex:big}, for readability. The full translation is
available in a companion preprint~\cite{bigraphs:long}.  Starting
from~\eqref{eq:ex:big}, a first step is to represent the place graph
more traditionally, i.e., as usual with trees.  But in order to avoid
confusion between the place and link graphs, we represent each node as
a cell, and adopt the convention that edges from the place graph
relate a principal port to a rightmost auxiliary port. Wires from the
link graph thus leave from other auxiliary ports.

\begin{figure}[t]
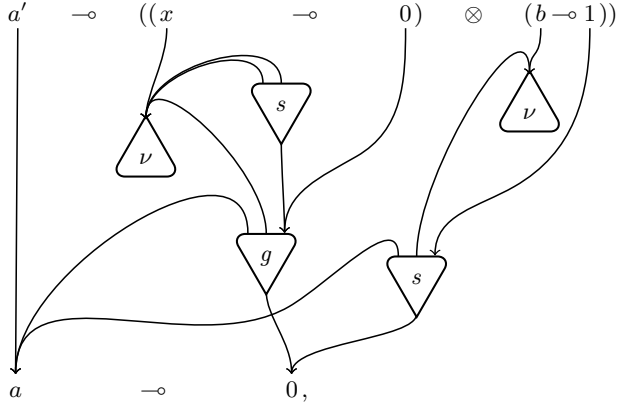

  \centering
  \begin{net}
    \matrix[column sep=0pt,at={(.6,4.5)}] (dom) { %
      \ssf{v}{a'} & \ssf{impl}{\hspace*{0.5cm}\impll\hspace*{0.5cm}} %
      & \lpare{} & \lpare{} & %
      \ssf{v'}{x} & \ssf{impldom}{\hspace*{1.5cm}\impll\hspace*{1cm}}%
      & \ssf{t}{0} & \rpare{} & %
      \ssf{tensdom}{\hspace*{.5cm}\tens\hspace*{.5cm}} %
      & \lpare{} & \ssf{v''}{b} & \gimpll{} & \ssf{t'}{1} & \rpare{} &
      \rpare{}\\}; %
    \matrix[column sep=0pt,at={(-1.5,-.5)}] (cod) { %
      \ssf{v'''}{a} &
      \ssf{impl}{\hspace*{1.5cm}\impll\hspace*{1.5cm}} %
      & \ssf{t''}{0} \\}; %
    \inetcell[at={(2,1)}](snd'){$\pisend$} %
    \inetcell[at={(0,1.3)}](get){$\piget$} %
    \inetcell[at={(.2,3.3)}](snd){$\pisend$} %
    \inetcell[at={(3.5,3.2)}](nu){$\nu$}[180] %
    \inetcell[at={(-1.6,2.6)}](nu'){$\nu$}[180] %
    \inetwire[<-](v'''.north)(v.south)%
    \inetwireext[<-](v'''.north) to[out=90,in=90] (get.right pax) ;%
    \inetwireext[<-] (v'''.north) to[out=90,in=210] (0.3,.6)
    to[out=30,in=90] (snd'.right pax) ; %
    \inetwireext[->](get.middle pax) to[out=90,in=90] (nu'.pal) ;%
    \inetwire[->](v'.south)(nu'.pal) %
    \inetwireext[->] (snd.right pax) to[out=90,in=90] (nu'.pal) ; %
    \inetwireext[->] (snd.middle pax) to[out=90,in=90] (nu'.pal) ; %
    \inetwire[->](snd.pal)(get.left pax) %
    \inetwireext[->] (t.south) to[out=-90,in=30] (1.2,2.6)
    to[out=210,in=90] (get.left pax) ; %
    \inetwire[->](snd'.pal)(t''.north) %
    \inetwire[->](v''.south)(nu.pal)%
    \inetwireext[->](snd'.middle pax) to[out=90,in=90] (nu.pal); %
    \inetwireext[->] (t'.south) to[out=-90,in=30] (3.3,2.2)
    to[out=210,in=90] (snd'.left pax) ; %
    \inetwire[->](get.pal)(t''.north) %
    \virgule{t''} %
  \end{net}
  \caption{A hybrid picture between bigraphs and proof nets}
\label{fig:hybrid}
\end{figure}
Finally, edges in $E$ in the bigraph are pointed to by ports and inner
names. We now represent them as (nullary) $\nu$ cells with pointers to
their principal port. We obtain the hybrid picture in
Figure~\ref{fig:hybrid}, where we have drawn the connectives to
emphasise the relationship with Figure~\ref{fig:main:ex}. And indeed
we have almost obtained the desired proof net. A first small problem
is the direction of wires in the linking graph which, intuitively, go
from occurrences of names to their creator (be it a $\nu$ or an outer
name). So we start by reversing the flow.

But that does not completely correct the mismatch, because in the case
of bound edges like $x'$ in our example bigraph, the $\nu$ cell is
absent in proof nets. But by Remark~\ref{rem:unique:binder}, the name
in question has a unique binding occurrence, and the $\nu$ cell may be
understood as an indirection between this binding occurrence and the
others. Contracting this indirection (and fixing the orientation
accordingly) yields exactly the desired proof net in
Figure~\ref{fig:main:ex}.

The procedure sketched on our example generalises, up to some
subtleties with unused names, and we have
\begin{theorem}
  This yields a functor $\bbig (\bsig) \rTo^{\T} S (\theory_\bsig),$
  which is faithful, essentially injective on objects, and neither
  full nor surjective on objects.
\end{theorem}
The functor is not strictly injective on objects, because two
isomorphic interfaces differing only by the choice of their set of
names have the same image under $\T$.

The functor $\T$ being non-full means that even between bigraphical
interfaces, $S (\theory_\bsig)$ contains morphisms which would be
ill-scoped according to Milner's scope rule. So it seems useful to
verify that the overall scoping discipline is maintained. This is
indeed the case, in the sense that $\T$ is full on whole programs,
i.e., bigraphs with no sites nor open names. Formally:
\begin{theorem}\label{thm:scope}
  The functor $\T$ induces an isomorphism on closed terms, i.e., an
  isomorphism of hom-sets $S (\theory_\bsig) (\un, t) \iso \bbig
  (\bsig) ((\emptyset, 0, \emptyset), (\emptyset, 1, \emptyset)).$
\end{theorem}
So, $S (\theory_\bsig)$ has as many whole programs as
$\bbig (\bsig)$, but more program fragments.

\section{Conclusions}

\paragraph{Related work}
Various flavours of closed categories have long been known to be
closely related to particular calculi with variable
binding~\cite{Lambek:categorical,Barber:action}. As mentioned in the
introduction, our approach may be considered as an update and further
investigation of Coccia et al.~\cite{Coccia}.  We should also mention
Tanaka's work on variable binding in a linear
setting~\cite{DBLP:conf/mfcs/Tanaka00}, whose relation to the present
work remains unclear to us.

A number of papers have been devoted to better understanding bigraphs,
be it as sortings~\cite{Debois:phd}, as cospans over
graphs~\cite{Sobocinski:graphslics}, as a compact closed
category~\cite{grohmann}, or as a language with variable
binding~\cite{Birkedal}. We appear to be the first to reconcile a full
treatment of scope (Theorem~\ref{thm:scope}) with algebraic tools,
i.e., seeing bigraphs as satisfying a universal property.

\paragraph{Future work}
We should try to push our approach further, e.g., by trying to use it
in an actual implementation. Also, we here only handle abstract
bigraphs, which do not have so-called \emph{relative pushouts}
(RPO). We thus should generalise our approach to deal with
\emph{concrete} bigraphs, be it in the form of Milner's original
pre-category or of Sassone and Soboci\'{n}ski's
\emph{G-categories}~\cite{Sobocinski:grpos}, and then try to construct
the needed (G)RPOs.

Another natural research direction from this paper concerns the
dynamics of bigraphs.  Our hope is that Bruni et
al.'s~\cite{Bruni:ccdc} very modular approach to dynamics may be
revived, and work better with \smc\ structure than with cartesian
closed structure. Specifically, with \smc\ structure, there is no
duplication at the static level, which might simplify matters.

\bibsettings{\tinybibsettings}
\bibliographystyle{jbwc}
\bibliography{bib}

\end{document}